\documentclass[utf8]{frontiersFPHY_mod} 

\setcitestyle{square} 

\usepackage{url,hyperref,lineno,microtype,subcaption}
\usepackage[onehalfspacing]{setspace}
\usepackage{float}
\usepackage{longtable}

\def\keyFont{\fontsize{8}{11}\helveticabold }

\def\firstAuthorLast{Narock {et~al.}} 
\def\Authors{Thomas Narock\,$^{1,*}$, 
            Ayris Narock\,$^{2, 3}$, 
            Luiz F. G. Dos Santos\,$^{4}$,
            and Teresa Nieves-Chinchilla\,$^{2}$}

\def\Address{$^{1}$Center for Data, Mathematical, \& Computational Sciences, \\ Goucher College, Baltimore, MD, USA \\
             $^{2}$ NASA Goddard Space Flight Center, Greenbelt, MD, USA \\
             $^{3}$ ADNET Systems Inc., Bethesda, MD, USA  \\
             $^{4}$CIRES, University of Colorado, Boulder, CO, USA}
\def\corrAuthor{Thomas Narock}
\def\corrEmail{Thomas.Narock@goucher.edu}

\begin{document}
\onecolumn
\firstpage{1}

\title[Flux Rope Orientation via Neural Networks]{Identification of Flux Rope Orientation via Neural Networks} 
\author[\firstAuthorLast ]{\Authors} 
\address{} 
\correspondance{} 

\extraAuth{}

\maketitle

Provisionally Accepted for Frontiers special issue: \\ Applications of Statistical Methods and Machine Learning in the Space Sciences \\
https://www.frontiersin.org/articles/10.3389/fspas.2022.838442/abstract \\

\begin{abstract}
\section{}
Geomagnetic disturbance forecasting is based on the identification of solar wind structures and accurate determination of their magnetic field orientation. For nowcasting activities, this is currently a tedious and manual process. Focusing on the main driver of geomagnetic disturbances, the twisted internal magnetic field of interplanetary coronal mass ejections (ICMEs), we explore a convolutional neural network's (CNN) ability to predict the embedded magnetic flux rope's orientation once it has been identified from in situ solar wind observations. Our work uses CNNs trained with magnetic field vectors from analytical flux rope data. The simulated flux ropes span many possible spacecraft trajectories and flux rope orientations. We train CNNs first with full duration flux ropes and then again with partial duration flux ropes. The former provides us with a baseline of how well CNNs can predict flux rope orientation while the latter provides insights into real-time forecasting by exploring how accuracy is affected by percentage of flux rope observed. The process of casting the physics problem as a machine learning problem is discussed as well as the impacts of different factors on prediction accuracy such as flux rope fluctuations and different neural network topologies. Finally, results from evaluating the trained network against observed ICMEs from Wind during 1995-2015 are presented.

\tiny
 \keyFont{ \section{Keywords:} flux rope, neural network, machine learning, space weather, magnetic field} 
 
\end{abstract}


\setlength{\parskip}{2pt}

\section{Introduction}
Coronal mass ejections (CMEs) are one of many manifestations of our dynamic Sun. CMEs are responsible for the transport of large quantities of solar mass into the interplanetary medium at very high speeds and in various directions. CMEs are commonly referred to as interplanetary coronal mass ejections (ICMEs) after leaving the solar atmosphere and reaching the interplanetary medium. ICMEs are the main drivers of geomagnetic activity at Earth as well as at other planets and spacecraft throughout the heliosphere \citep{baker_2008}\citep{Kilpua2017}. In situ observations of ICMEs frequently find them to have a combination of an increase in magnetic field strength, low proton plasma temperature, $\beta_{plasma}$ below 1, and monotonic rotation of the magnetic field components \citep{burlaga_1988}. These characteristics are commonly referred to as a Magnetic Cloud (MC) \citep{burlaga_1981}\citep{klein_burlaga_1982}. CME eruption theories \citep{Vourlidas_2014} suggest that a twisting internal magnetic signature - referred to as a flux rope - is always present. While commonly observed, not all ICMEs show the signatures of an internal structure characterized by a flux rope, perhaps resulting from changes during interplanetary evolution \citep{Jian2006}\citep{manchester_2017}. Yet, flux ropes are sufficiently prevalent that they can aid in space weather forecasting. The observed magnetic field profile depends on a flux rope's orientation and where the spacecraft traverses the structure. The latitudinal and longitudinal deflections of CMEs happen in the lower corona and are not expected to change greatly throughout the interplanetary medium. If flux rope orientation and the spacecraft's crossing trajectory can be determined early enough, this can lead to advanced forecasting as the remaining portion of the flux rope's magnetic field structure can be inferred from physics-based models. The flux rope's internal magnetic field structure is prone to couple with Earth's upper magnetosphere triggering magnetic reconnection processes and allowing the injection of solar magnetic energy into the magnetospheric system. Orientation determines the magnetic field profile observed at Earth and, thus, the geo-effectiveness of the flux rope making early determination of a flux rope's orientation a vital requisite for space weather forecasting. A major challenge to developing such a forecasting system is that information about the internal magnetic structure of ICMEs is often limited to 1D observations of a single spacecraft crossing the structure. This leaves a considerable amount of uncertainty about the three-dimensional structure of the ICME.  

Various physics-based flux
rope models exist (for example \citet{Lepping_Jones_Burlaga_1990} and \citet{Nieves_2019}) that can be used to reconstruct the internal
ICME magnetic configuration and provide information on orientation, geometry,
and other magnetic parameters such as the central magnetic field. Recent in situ observations \citep[][and references therein]{Nieves_2018, Nieves_2019, rodriguez_2021, kilpua_2017} are continuing to complement earlier studies \citep{gosling_1973, burlaga_1981, klein_burlaga_1982} and enhance our understanding of ICMEs, MCs, and flux ropes. Meanwhile, an increase of space- and ground-based data availability has led to more interest in applications of machine learning within the space weather community \citep[see][and references therein]{Camporeale_2019}. Nguyen et al. \citep{Nguyen2018} have explored machine learning techniques for automated identification of ICMEs and \citet{dossantos_2020} used a deep neural network to create a binary classifier for flux ropes in the solar wind, determining whether a flux rope was or was not present in a given interval. Recently, Reiss et al. \citep{Reiss2021} use machine learning to predict the minimum Bz value as a magnetic cloud was sweeping past a spacecraft.

We aim to assess a neural network's ability to predict a flux rope's orientation after an ICME is identified. This work is an attempt to understand if a neural network can predict a flux rope's orientation having only seen a portion of the event. If the full magnetic field profile of the flux rope can reliably be reconstructed when the spacecraft is only partially through the flux rope this can provide advanced warning of impending geomagnetic disturbance. Yet, as machine learning is relatively new to space weather, the accuracy of these forecasts, and more generally, which neural network topologies to utilize, are unclear. We begin with a set of exploratory experiments to quantify the capabilities of neural networks in this regard. The results of these experiments then serve as a baseline to begin exploring forecasting. 

Here, we extend the binary classifier work of \citet{dossantos_2020} and explore a neural network’s ability to predict the orientation, impact parameter, and chirality of an already identified flux rope. We extend the capabilities presented in Reiss \citep{Reiss2021} by reconstructing the entire three dimensional magnetic field profile. The neural network is trained using simulated magnetic field measurements over a range of spacecraft trajectories and flux rope orientations. Moreover, we report on the prediction accuracy of the neural network as a function of percentage of flux rope observed. To connect this proof of concept to its potential for real-world use, we also present results from evaluating the neural network on flux ropes observed by the Wind spacecraft. In performing these experiments, we highlight the multiple ways in which this space weather forecasting problem can be cast as a machine learning application and the implications those choices have on prediction accuracy. 

In section 2 we present our methodology. We describe the flux rope analytical model and the generation of our synthetic data set. Section 2 also details our neural network designs and training process. Section 3 presents our results first from the full duration synthetic flux ropes, then from partial duration flux ropes, and ultimately from application to flux ropes observed from the Wind spacecraft. We present a discussion of these results in section 4 along with concluding remarks.

\section{Methodology}
The task of predicting a flux rope’s key defining parameters from magnetic field measurements can be cast as a supervised machine learning problem. This is an approach in which the goal is to learn a function that maps an input to an output based on numerous input-output pairs. There are currently not enough in situ observed flux ropes (inputs) with known key parameters (outputs) to train a neural network. Instead, we choose to use a physics-based flux rope model to produce a synthetic training dataset.

\subsection{Synthetic Data}
The circular-cylindrical flux rope model of \citet{Nieves_C_2016} (N-C model) is used to simulate the magnetic field signature of flux ropes at numerous orientations and spacecraft trajectories. The N-C model takes as input the following parameters:
\begin{longtable}{l p{10cm}}
    $H$ & Chirality of the flux rope; Right-handedness is designated with 1, left-handedness with -1.\\
    \\
    $Y_0$ & Impact parameter; The perpendicular distance from the center of the flux rope to the crossing of the spacecraft expressed as a percentage of the flux rope's radius.\\
    \\
    $\phi$ & Longitude orientation angle of the flux rope.\\
    \\
    $\theta$ & Latitude orientation angle of the flux rope. \\
    \\
    $R$ & Radius of flux rope\\
    \\
    $V_{sw}$ & Bulk velocity of the solar wind\\
    \\
    $C_{10}$ & A measure of the force free structure. A value of 1 indicates a force free flux rope.
    

\end{longtable}
\addtocounter{table}{-1}

\begin{figure}[H]
    \begin{center}
    \includegraphics[width=\textwidth]{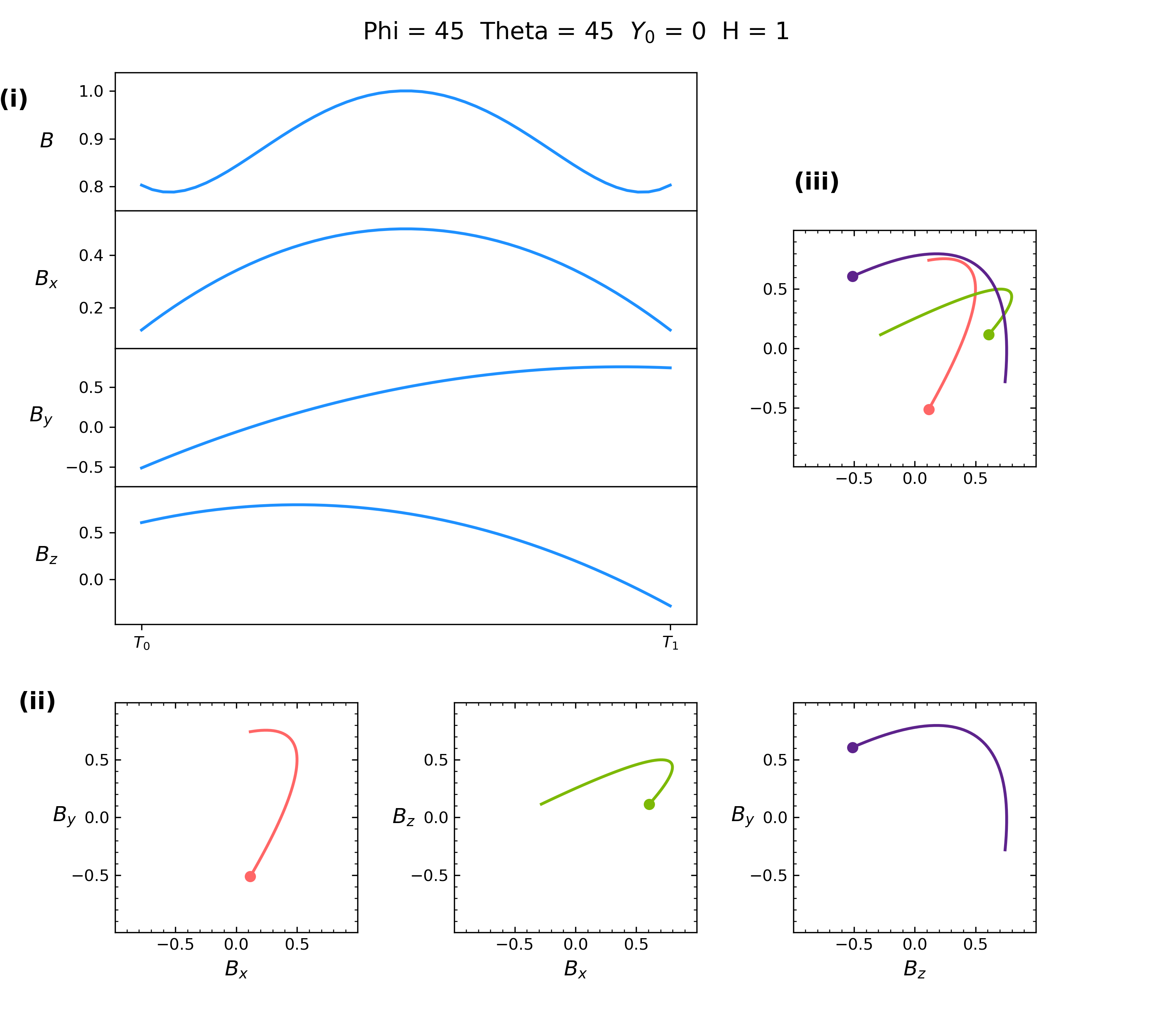}
    \end{center}
    \caption{A synthetic flux rope example generated using $\phi = 45$, $\theta = 45$, $Y_0 = 0$, and $H = +1$. (i) The total magnetic field and the magnetic field components. (ii) Three hodograms of the magnetic field components. The dot represents the starting point of the simulated flux rope crossing.  Flux rope classification with 2D CNNs require three images, which can (iii) be combined for a single set of convolutions or (ii) have convolutions applied separately.}
    \label{fig:synthetic_data}
\end{figure}

The output of the N-C model is the magnetic field profile (Bx, By, Bz components) that would be observed for spacecraft traversing a flux rope with the given input parameters. An illustration of this is shown in Figure \ref{fig:synthetic_data} where panel (i) shows the N-C model output visualized as a time series and panel (ii) depicts the same output as hodograms. All flux ropes were simulated using a solar wind speed ($V_{sw}$) of $450 km/s$, a radius ($R$) of $0.07 AU$, and with poloidal normalization. The $C_{10}$ parameter was held constant at 1, which imposes a force free structure. The model was run for all combinations of longitude ($\phi$) $\in [5^\circ, 355^\circ]$, latitude ($\theta$) $\in [-85^\circ, 85^\circ]$, and impact parameter ($Y_0$) $\in [0\%, 95\%]$. This is done in $5^\circ$ and $5\%$ increments and with both chirality options, $H \in \{-1, 1\}$. We exclude combinations involving $\phi = 180^\circ$ as the model is not always defined in this instance.  This results in 98,000 combinations. 

The fixed bulk velocity of $450 km/s$ and fixed radius of $0.07 AU$ describe a ``typical'' flux rope observed at Earth based on fittings in \citep{Nieves_2019}. Magnetic field profiles of this ``typical'' flux rope have been shown \citep{dossantos_2020} to scale with changes in speed and size. In other words, magnetic field profiles are very similar when orientation is held constant and speed and radius are varied. The only variation in the profiles is duration, which is not a factor for us as all flux ropes are interpolated to 50 points. This relationship allows us to only simulate a subset of all possible speeds and sizes drastically reducing the training data set size and minimizing training time.

The output from each of these $98,000$ combinations is then used to generate 10 exemplars of this event in different percentages of completion - from $10$\% to $100$\% in steps of $10$\%. For example, first a 50-point trace through a flux rope defined by the parameter combination is generated ($100\%$ completion, Figure \ref{fig:synthetic_data}(i)).  The first $5$ points are interpolated to $50$ points to create the $10\%$ completion exemplar. Similarly the first $10$ points are used to create the $20\%$ exemplar, the first $15$ points for the $30\%$ exemplar, etc. The final dataset contains $980,000$ exemplars - a mixture of full duration and partially observed events.  These simulated partial flux ropes are useful to understand how much of the flux rope needs to be observed before reliable autonomous predictions can be made. The ability to predict in the absence of the complete flux rope is very desirable in the context of space weather forecasting.

\subsection{Convolutional Neural Networks}
Simply put, a convolution is the application of a filter to an input that results in an activation. Repeatedly applying the same filter to an input – for example, by sliding a small dimensional filter across an image - results in a map of activations called a feature map. The feature map then indicates the locations and strength of a detected feature in the input.  Convolutions are the major building blocks of convolutional neural networks (CNNs) \citep{lecun1995}, which use a training dataset to learn a set of highly specific filters from the input that lead to the most accurate output predictions. The innovation of the CNN is in not having to handcraft the filters, but rather automatically learning the optimal set of filters during the training process.

The CNN is the basis of the neural network architectures explored in this work. The training phase consists of showing the network the input-output pairs of simulated flux rope magnetic field vectors (input) and the corresponding key parameters used to create this simulated data trace (output). The key parameters represented in this training are $\phi$, $\theta$, $Y_0$, and $H$. From repeated exposure to input-output pairs the network learns the filters that lead to the most optimal predictions. These neural networks require all inputs to be of the same size, which does not pose a problem when working with synthetic data. In situ observations from spacecraft, however, reveal a diverse set of events ranging from a few hours to multiple days. These need to be thoughtfully processed for use as input to the CNN. One could average or interpolate in situ events to ensure all input magnetic field time series are of the same length. Alternatively, \citet{dossantos_2020} showed an innovative technique of representing flux ropes as hodograms. Flux ropes of any duration can be cast as a set of three consistently sized images (see Figure \ref{fig:synthetic_data}(ii)), which can then serve as input to a CNN. This technique also leverages a wide swath of existing literature in the computer vision field (particularly in the area of handwritten digit classification) that can be helpful in fine-tuning the CNN architecture. Over the next several sections, we present a series of experiments evaluating multiple CNN architectures. Specifically, we compare the predictions from convolutions applied directly to magnetic field time series to predictions made from convolutions applied to hodograms of those magnetic field time series. We do so under two scenarios. First, we develop a baseline for a CNN's capability to predict flux rope orientation by training the architectures with only the 98,000 exemplars of full duration flux ropes. We separately train another copy of each of the three aforementioned architectures with the complete set of 980,000 full and partial duration flux ropes to assess CNN usage in a time-predictive capacity.

\subsubsection{CNN Architectures}
Representing flux ropes as hodograms was inspired by work in handwritten digit classification \citep{dossantos_2020}. Yet, flux ropes provide a more challenging version of this computer vision problem. The input for handwritten digit classification is always a single image; however, flux ropes require a set of three images (hodograms) to capture the entirety of their magnetic field configurations. An initial research question is then how to feed three hodograms as input to a CNN. In the approach chosen for our first architecture, we stack the images (Figure \ref{fig:synthetic_data}(iii)) and do a single two-dimensional convolution across the resulting tensor. In our second tested architecture, we apply two-dimensional convolutions to each of the three hodograms separately (Figure \ref{fig:synthetic_data}(ii)) and then concatenate the resulting feature maps. 

\begin{figure}[H]
  \begin{center}
  \includegraphics[width=\textwidth]{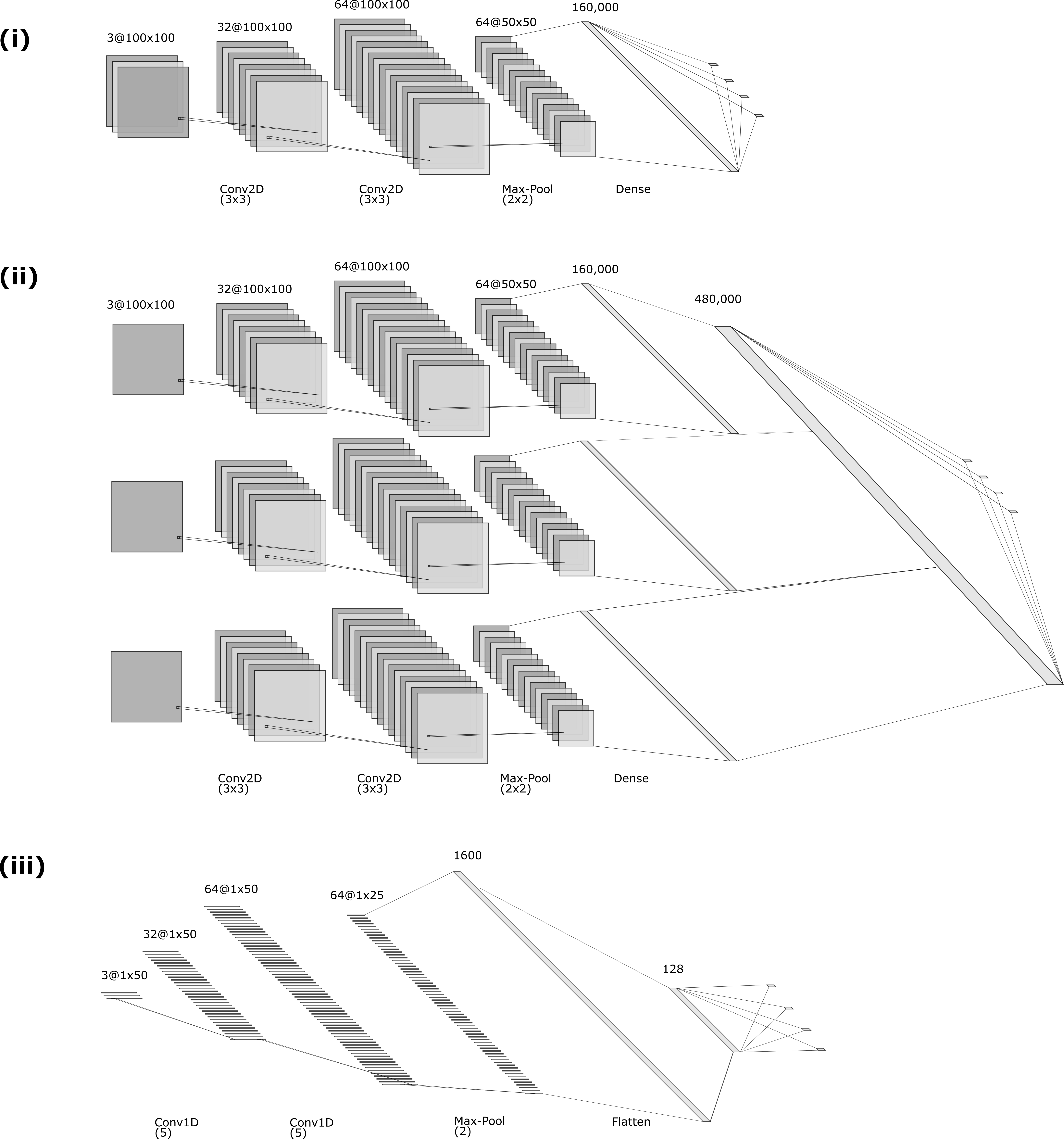}
  \end{center}
  \caption{CNN architecture schematics. (i) 2D CNN with one input which uses stacked hodograms; (ii) 2D CNN with three inputs, which performs individual convolutions over each of the three hodograms, and (iii) CNN architecture for 1D convolutions over time series}
  \label{fig:CNN_Arch}
\end{figure}

The architecture schematic for the stacked approach is shown in Figure \ref{fig:CNN_Arch}(i). An input layer of dimension $[100,100,3]$ passes through two rounds of 2D Convolution with a $3\times3$ kernel size. The resulting layer of dimension $[100,100,64]$ undergoes a $2\times2$ Max Pooling to transform to dimensions $[50, 50, 64]$.  This layer is then Flattened and Fully Connected to each of four output layers. This 2D CNN with one input ends up with 979,398 trainable parameters.

The architecture for applying two-dimensional convolutions to each of the three hodograms separately and then concatenating the resulting feature maps is shown in Figure \ref{fig:CNN_Arch}(ii). Each prong of the initial part of this network involves the same transformations as in the previously described network, with the exception that each of the three input layers is of dimension $[100, 100, 1]$. Additionally, the Flattened layers at the end of these individual pipelines are then concatenated before being Fully Connected to the four output layers. This architecture has the advantage that salient features in specific hodograms can become more apparent in the feature maps. Yet, this comes at the cost of a more complex neural network. With 2,936,454 trainable parameters, this CNN has significantly more weights that need training.

Finally, we tested an architecture that did not rely on hodogram images. Instead we apply 1D convolutions directly to magnetic field time series. This approach is depicted in Figure \ref{fig:CNN_Arch}(iii) and results in the smallest CNN with 216,518 trainable parameters. The input layer of size $[1, 50, 3]$ has a 1D Convolution with kernel size $5$ applied twice, resulting in a layer of dimension $[1, 50, 64]$. Max Pooling with a kernel size $2$ then creates a layer of dimension $[1, 25, 64]$ before this is flattened to a vector of size $1600$. This layer is then Fully Connected to a layer of size $128$ and then to each of the four output layers.

Hyperparameters for all of these architectures were found by doing a simple grid search. Our focus was on comparison of architectures and we acknowledge there may still be room for hyperparameter optimization. 

\subsubsection{CNN Tuning and Training}
Neural networks learn by minimizing a loss function, which typically involves some measure of difference between current predictions and expected outputs. Angles can challenge neural network predictions in that loss functions, such as mean squared error (MSE), completely miss the circular nature of angles. For example, if a flux rope's longitudinal value is $0^{\circ}$, then predictions of $350^{\circ}$ and $10^{\circ}$ are both off by $10^{\circ}$. Yet, MSE will miss this relation and penalizes the $350^{\circ}$ prediction more than the $10^{\circ}$ prediction. To combat this, we predict $(\sin(\measuredangle), \cos(\measuredangle))$ with $\tanh$ activation to enforce outputs to be in $[-1, +1]$. We then post-process the CNN's predictions with $\arctan$ to convert to degrees. This approach is applied across all three CNN architectures when predicting $\phi$ and $\theta$.

A challenge also arises in that predicting the real-valued parameters $\phi$, $\theta$, and $Y_0$ is a regression problem while determining the binary parameter, $H$, is a classification problem. We address this by training four separate loss functions in each CNN. For $\phi$ and $\theta$ we predict the pair $(\sin(\measuredangle), \cos(\measuredangle))$ and train using the MSE loss function. Impact parameter is also trained using MSE while chirality is defined as a two class classification problem and trained using binary cross entropy. 

In our first experiment, the 98,000 full duration synthetic flux ropes were randomly divided into $60\%$ training, $20\%$ validation, and $20\%$ testing sets. This resulted in $58,800$ synthetic flux ropes used for training, 19,600 used for validation, and $19,600$ used for testing. The training set was used in a supervised learning fashion with the Adam optimizer \citep{kingma_2015} with the validation set used during the training process to avoid overfitting. All networks were set to train over $500$ epochs, but the 2D CNNs had early stopping from criteria on the validation set at around $35$ to $50$ epochs. The 1D CNN had a training time of 12 minutes and both the one input and three input 2D CNNs had training times approaching $4$ to $6$ hours. 

The setup of the second experiment, in which we train over all full and partial flux ropes, was similar. A $60/20/20$ split was used, with validation criteria used for early stopping and evaluation on the testing set. Again, the 1D CNN trained over all $500$ epochs while the 2D networks reached early stopping within $50$ epochs. The 1D CNN took just over 2 hours to train, while the 2D CNNs completed in $6$ to $10$ hours. It should be noted that all percentages of a particular flux rope configuration were included in an input batch. Also, an important consideration in this scenario is that the networks will be seeing multiple inputs that share the same output. All neural networks were constructed, trained, and tested using Python 3.8.10, Keras 2.4.3 \citep{keras}, TensorFlow 2.3.1 \citep{tensorflow}, Numpy 1.18.5 \citep{numpy}, and Scipy 1.7.1 \citep{scipy}.

\subsection{Wind spacecraft}\label{Wind_data}
The final segment of this work is to evaluate the trained CNNs on flux ropes observed by the Wind spacecraft. This application of the CNNs on non-synthetic data helps us understand the limitations of the flux rope analytical model and the transition to actual space weather forecasting. 
\citet{Nieves_2018} carried out a comprehensive study of the internal magnetic field configurations of ICMEs observed by Wind at $1 AU$ in the period 1995-2015. In this analysis, the term magnetic obstacle (MO) is adopted as a more general term than magnetic cloud in describing the magnetic structure embedded in an ICME. The authors used the Magnetic Field Instrument (MFI) \citep{Lepping_1995} and Solar Wind Experiment (SWE) \citep{Ogilvie_1995} to manually set the boundaries of the MO through visual inspection. All MO events were sorted into three broad categories based on the magnetic field rotation pattern: events without evident rotation (E), those with single magnetic field rotation (F), and those with more than one magnetic field rotation (Cx). More recently, \citet{Nieves_2019} presented an in-depth classification, which further classified the F types events into F-, Fr, and F+ based on the angular span of the magnetic field rotation. These events were then manually fit with the Circular-Cylindrical N-C model by a human expert. Of the events cataloged and fit, those that were classified as the Fr type tended to be the ones that could best be fit with the N-C model. Because we restricted out training set of synthetic data to flux rope cases with a $Y_0 > 0$, we also restrict our Wind test event cases to this criteria.  We use this subset of 75 Wind Fr type events to evaluate our neural network predictions on actual flux rope observations.  We compare the human-fit key parameters to the neural network predictions. While we have high confidence in the human expert's fit values, we acknowledge that they are not definitive. Other experts may parameterize the event slightly differently. Instead of using the human expert as ground-truth, we are interested in seeing if a neural network, trained on the same physical model that the human expert used, will arrive at similar flux rope orientations. The average correlation coefficient is used to compare human and neural network fits to the Wind magnetic field profiles.

As noted earlier, the 1D CNN is configured to input vectors of size $50$ and trained on normalized synthetic data, requiring some pre-processing for use with real-event data. We begin with the 1-minute resolution MFI data for each the $75$ Wind events and apply a 5-point moving average smoothing followed by interpolation to $50$ points evenly spaced in time.

\section{Results}
\subsection{Full Duration Synthetic Flux Ropes}
\begin{figure}[H]
  \begin{center}
  \includegraphics[width=\textwidth]{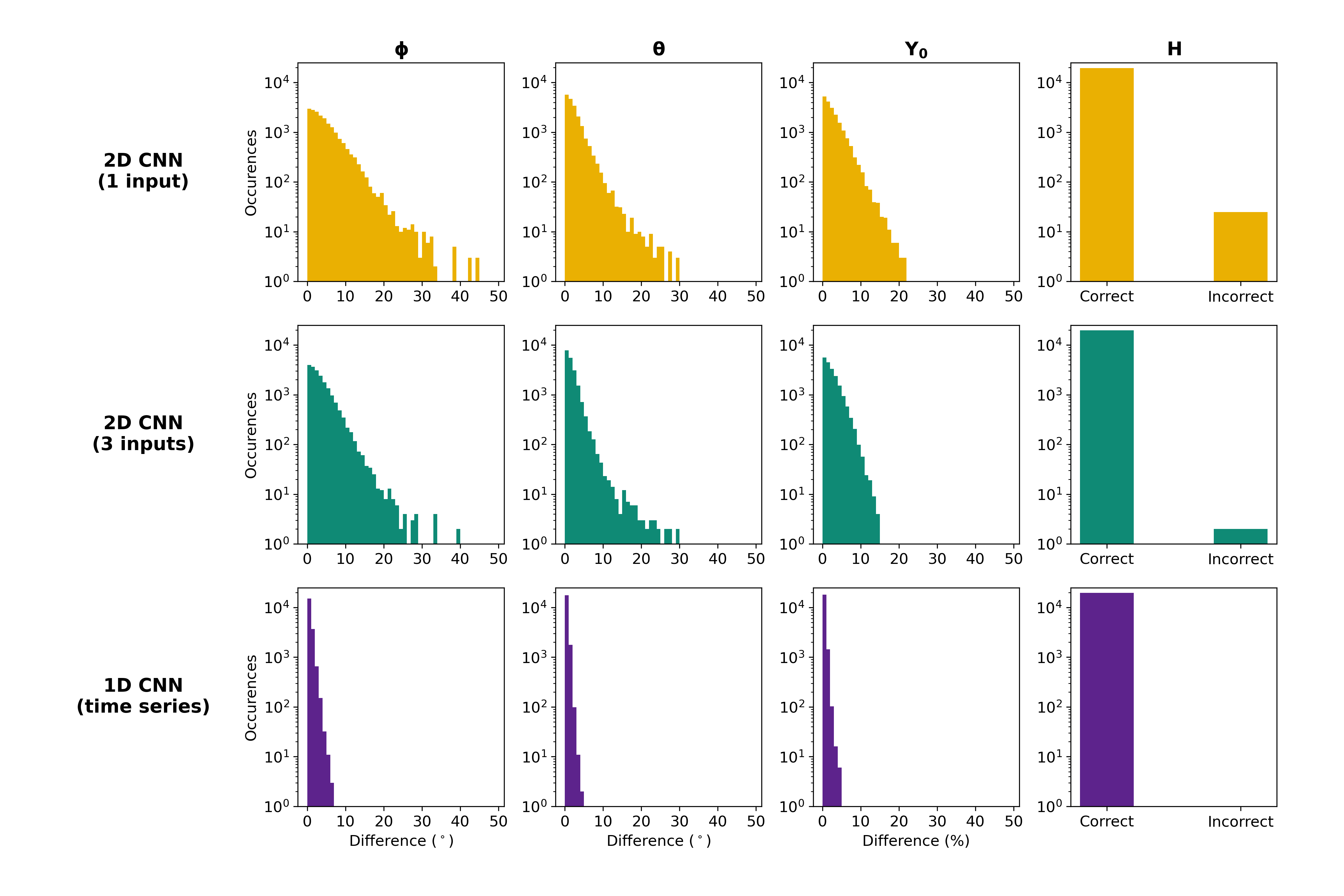}
  \end{center}
  \caption{The parameter prediction error for the synthetic test set of full duration flux rope crossings.  The first three columns show histograms of the differences between predicted and modeled $\phi$, $\theta$, and $Y_0$ values.  The last column displays the number of correct and incorrect chirality predictions.  The 1D CNN predicted all angles within $10^\circ$, all $Y_0$ within $10\%$ and achieved $100\%$ accuracy for $H$. While the 2D CNNs perform this well in most cases, they exhibit a much wider range in error.}
  \label{fig:Results_1_full}
\end{figure}

Results of applying the neural networks trained on full duration flux ropes to the testing set of full duration flux ropes are shown in Figure \ref{fig:Results_1_full}. The $\phi$, $\theta$,  and $Y_0$ panels display histograms of the difference between the neural network’s predictions and the true values used to create the simulated instance for longitude, latitude, and impact parameter, respectively. The $H$ panel shows the number of correct and incorrect chirality predictions. Subsequent figures use the same color scheme (2D CNN with 1 input in orange, 2D CNN with 3 inputs in green, and 1D CNN in purple) for clarity. Table \ref{table:full_stats} lists the median values of these difference distributions.  The 2D CNN with a single input channel has the highest median difference across all three of the real-valued key parameters, as well as the most skewness. The 1D CNN shows the least skewness and the lowest median difference values across the parameters.  The 2D CNN with three inputs falls in between, but with median difference and skewness more similar to the other 2D network than the 1D.  A similar trend is seen in the $H$ predictions, with the one input 2D network having the most incorrect classifications and the 1D network making no incorrect classifications.  Taken together, it is evident that the 1D CNN, which is applied to the time series directly, gives more accurate predictions across all four output parameters. 

\begin{table}[H]
    \centering
    \begin{tabular}{c|c|c|c}
        \multicolumn{4}{c}{Median Difference}\\ \hline
        CNN & 2D (1) &  2D (3) & 1D \\ \hline
        $\phi(^\circ)$   & 3.65 & 2.67 & 0.54 \\
        $\theta(^\circ)$ & 1.86 & 1.31 & 0.37 \\
        $Y_0$ (\%)       & 2.13 & 1.93 & 0.34 \\
    \end{tabular}
    \caption{Median of the parameter differences shown in Figure  \ref{fig:Results_1_full}. }
    \label{table:full_stats}
\end{table}

\begin{figure}[H]
  \begin{center}
  \includegraphics[width=\textwidth]{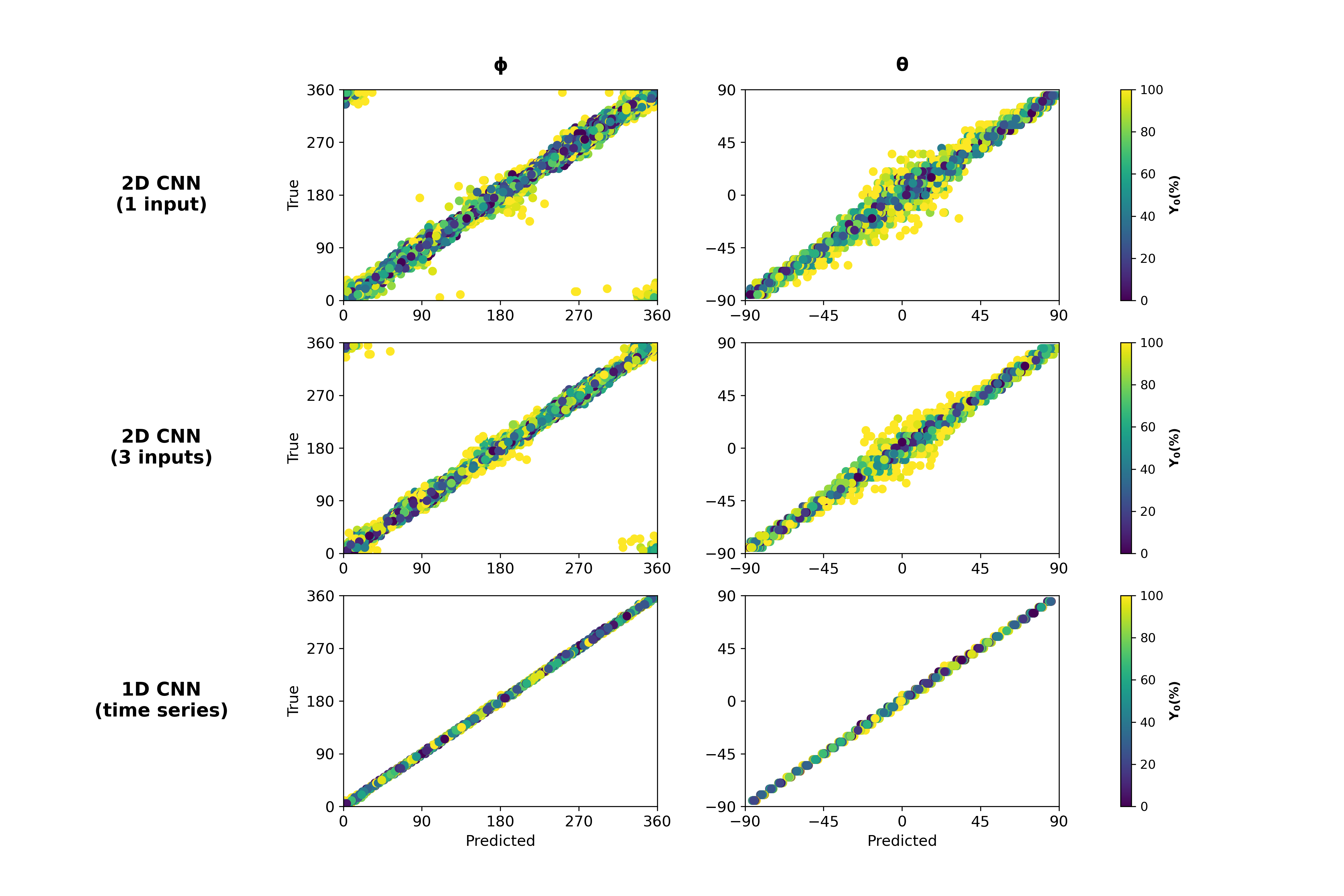}
  \end{center}  
  \caption{Latitude and longitude predictions vs. true values as a function of spacecraft impact parameter when evaluated on synthetic data test set. The 1D CNN performs similarly well across the entire range of $Y_0$ while the 2D CNNs show a larger discrepancy in parameter predictions at high impact parameters of $Y_0 > 80\%$.}
  \label{fig:Results_2_full}
\end{figure}

While the 1D CNN gives the most accurate predictions, all three architectures give reasonably useful predictions for the vast majority of cases.  The bulk of the prediction errors are less than $15^\circ$ for $\phi$ and $\theta$ and under $10\%$ for $Y_0$ for both of the 2D CNNs.  Figure \ref{fig:Results_2_full} illustrates the prediction errors as a function of $Y_0$.  The 2D CNNs using hodograms as input have the most significant $\phi$ and $\theta$ prediction errors, which occur at large $Y_0$.  In contrast, the 1D CNN more accurately predicts $\phi$ and $\theta$ over the entire range of simulated $Y_0$. Clearly, the architecture of the neural network plays a role in prediction accuracy and leads to an important trade off. The two-dimensional networks, by using hodogram input, remove time from the training process.  This makes little difference with the synthetic training data but is an advantage when working with data from time-varying, real ICME events, as the data can be used with less manipulation in pre-processing.  Yet, this comes at the cost of less accurate predictions at large spacecraft impact parameters ($Y_0$). The trade off is that the simpler and more accurate 1D network comes with the added complexity of determining the most appropriate data transformations to fit the measured time-series to the prescribed input array dimensions of the network.

Our CNNs were each designed with four loss functions and our analysis up to this point has looked at each predicted parameter individually. We now turn our attention to evaluating the predictions as a set. To do so, we use the predicted $\phi$, $\theta$, $Y_0$, and $H$ to reconstruct the magnetic field time series with the analytical model and correlate it with the simulated magnetic field used as input for the CNN.  For analysis, we use the average correlation coefficient, $r$, defined as:

\begin{equation}
    r = \frac{ r_x + r_y + r_z }{3}
\end{equation}
where $r_x$ is the Pearson's correlation between the simulated and reconstructed $b_x$ components, $r_y$ is the Pearson's correlation between the simulated and reconstructed $b_y$ components, and $r_z$ is the Pearson's correlation between the simulated and reconstructed $b_z$ components.

\begin{figure}[H]
  \begin{center}
  \includegraphics[width=\textwidth]{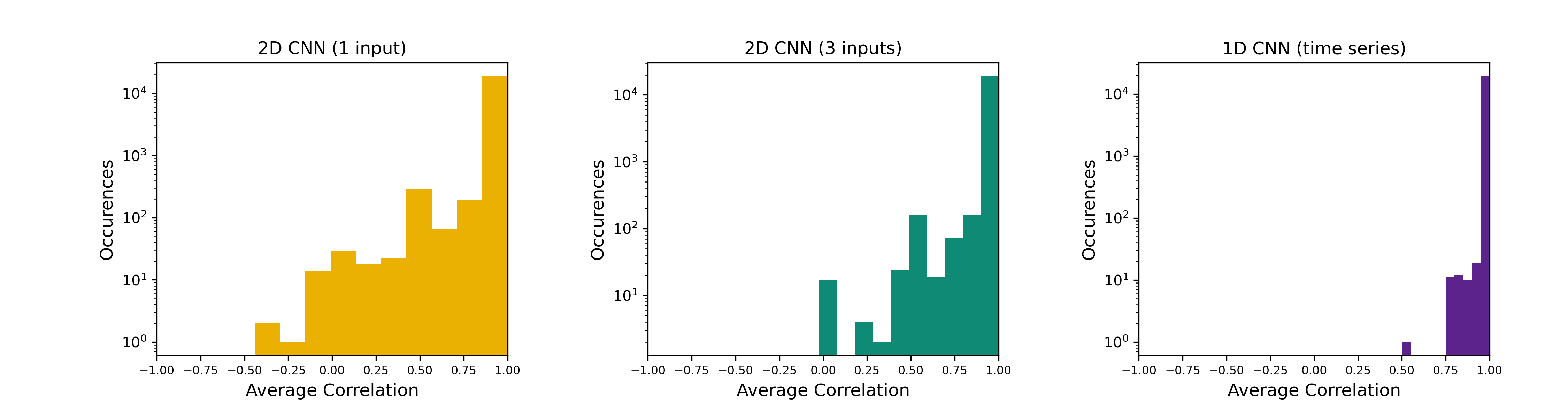}
  \end{center}
  \caption{Correlation coefficient histograms on full duration, synthetic data test set for each neural network architecture. Each set of parameters $\{\phi, \theta, Y_0, H\}$ model a spacecraft's traversal of a flux rope.  In these comparisons, the magnetic field trace modeled by the predicted parameters is correlated with the trace modeled by true parameters. Again, we see that all three architectures predict highly correlated results in the vast majority of cases but with the 2D CNNs exhibiting a significantly wider distribution.}
  \label{fig:Results_3_full}
\end{figure}

Figure \ref{fig:Results_3_full} shows the average correlation values for each of the three CNN architectures.  While the bulk of the simulated data predicted by the 2D CNNs have high average correlation, over 0.75, there is a long tail of predictions with much lower correlation.  The single input 2D CNN even makes some predictions that lead to negative correlations.  As in the individual key parameters, we see the 1D CNN applied to the time series outperforming the 2D CNNs. In the case of the 1D CNN, we find only one correlation value below 0.75. Further analysis reveals that this event occurred at a simulated $\phi$ value of $175^{\circ}$. The 1D neural network predicted a $\phi$ value of $181^{\circ}$. Although the neural network predictions were fairly accurate (within $5^{\circ}$, 5\%, and correct $H$), this small deviation in $\phi$ changed the spacecraft's trajectory through the flux rope leading to a negative correlation in the $b_z$ component. Because the 2D CNNs are impacted by their difficulties making predictions at large spacecraft impact parameters, we see many of the poor average correlation coefficients in the 2D CNNs at large spacecraft impact parameters.

\subsection{Partial Duration Synthetic Flux Ropes}
In our second experiment, we retrained a second version of each of the three neural networks, this time using the full set of 980,000 full and partial duration flux ropes. Like the difference comparisons shown in Figure \ref{fig:Results_1_full} and Table \ref{table:full_stats}, Table \ref{table:partial_stats} provides summary statistics of $\phi$, $\theta$, and $Y_0$ prediction error as a function of percentage of flux rope observed. All three models make fairly accurate predictions even when seeing just $10\%$ of the flux rope and then continue to improve their prediction accuracy up to a point. After this point, the key parameter accuracy gets worse as higher percentages of the flux ropes are fed to the networks.  The level of observation giving the lowest median errors for each CNN is highlighted in yellow, with the next lowest medians highlighted in green. Additionally, all three models were able to predict the correct $H$ over $99\%$ of the time at all percentages of flux rope observed.

\begin{table}[H]
    \centering
    \begin{tabular}{|c|c|c|c|||c|c|c|||c|c|c|}
    \multicolumn{1}{c}{} & \multicolumn{3}{c}{2D (1)} &  \multicolumn{3}{c}{2D (3)} & \multicolumn{3}{c}{1D} \\
    \hline
    \% Observed & $\phi$ & $\theta$ & $Y_0$ & $\phi$ & $\theta$ & $Y_0$ & $\phi$ & $\theta$ & $Y_0$ \\
    \hline
    10  & $7.67^{\circ}$ & $5.85^{\circ}$ & 7.33\% & $5.49^{\circ}$ & $3.92^{\circ}$ & 5.61\% & $2.10^{\circ}$ & $1.23^{\circ}$ & 1.28\% \\ \hline
    20  & $6.62^{\circ}$ & $4.89^{\circ}$ & 6.62\% & $4.94^{\circ}$ & $3.47^{\circ}$ & 5.15\% & $1.58^{\circ}$ & $1.02^{\circ}$ & 1.08\% \\ \hline
    30  & $6.25^{\circ}$ & $4.58^{\circ}$ & 6.20\% & \cellcolor{olive}$4.70^{\circ}$ & \cellcolor{olive}$3.28^{\circ}$ & 4.98\% & $1.37^{\circ}$ & $0.90^{\circ}$ & 0.94\% \\ \hline
    40  & $6.29^{\circ}$ & $4.38^{\circ}$ & 5.99\% & \cellcolor{yellow}$4.60^{\circ}$ & \cellcolor{yellow}$3.24^{\circ}$ & \cellcolor{olive}4.96\% & $1.23^{\circ}$ & $0.84^{\circ}$ & 0.84\% \\ \hline
    50  & \cellcolor{yellow}$5.96^{\circ}$ & \cellcolor{olive}$4.28^{\circ}$ & \cellcolor{olive}5.87\% & \cellcolor{olive}$4.61^{\circ}$ & \cellcolor{yellow}$3.24^{\circ}$ & \cellcolor{yellow}4.95\% & $1.14^{\circ}$ & $0.81^{\circ}$ & 0.79\% \\ \hline
    60  & \cellcolor{olive}$6.02^{\circ}$ & \cellcolor{yellow}$4.22^{\circ}$ & \cellcolor{yellow}5.75\% & \cellcolor{olive}$4.70^{\circ}$ & \cellcolor{olive}$3.31^{\circ}$ & \cellcolor{olive}4.96\% & $1.11^{\circ}$ & $0.80^{\circ}$ & 0.76\% \\ \hline
    70  & $6.17^{\circ}$ & \cellcolor{olive}$4.23^{\circ}$ & \cellcolor{olive}5.79\% & $4.74^{\circ}$ & $3.33^{\circ}$ & 5.06\% & \cellcolor{olive}$1.04^{\circ}$ & \cellcolor{olive}$0.76^{\circ}$ & \cellcolor{olive}0.72\% \\ \hline
    80  & \cellcolor{olive}$6.13^{\circ}$ & $4.36^{\circ}$ & 5.89\% & $4.83^{\circ}$ & $3.37^{\circ}$ & 5.10\% & \cellcolor{yellow}$1.01^{\circ}$ & \cellcolor{yellow}$0.75^{\circ}$ & \cellcolor{yellow}0.71\% \\ \hline
    90  & $6.38^{\circ}$ & $4.45^{\circ}$ & 6.13\% & $4.93^{\circ}$ & $3.41^{\circ}$ & 5.25\% & \cellcolor{olive}$1.04^{\circ}$ & \cellcolor{olive}$0.76^{\circ}$ & \cellcolor{olive}0.75\% \\ \hline
    100 & $7.07^{\circ}$ & $4.81^{\circ}$ & 6.52\% & $5.17^{\circ}$ & $3.61^{\circ}$ & 5.51\% & $1.10^{\circ}$ & $0.79^{\circ}$ & 0.83\% \\ \hline
    \end{tabular}
    \caption{Median parameter differences by percentage of flux rope observed for the neural network architectures when trained using partial duration crossings. Cells highlighted in yellow indicate the lowest error for each (CNN, parameter) pair and cells highlighted in green, the next two lowest errors. The overall performance of the 1D CNN continues to be significantly better than the 2D CNNs. The 2D CNNs make their best predictions when seeing less of the flux rope crossing.}
    \label{table:partial_stats}
\end{table}

It is worth noting that all three networks perform worse at $100\%$ duration when trained with partial duration flux ropes as compared to these same networks trained only with full duration flux ropes. The introduction of partial flux ropes into the training produces more error (see Table \ref{table:full_stats} and Table \ref{table:partial_stats}). We suspect this is due to multiple inputs now producing the same output. It remains for future research to conduct a more in depth analysis into how to combat this.

As with the networks trained only with full duration flux ropes, the 1D CNN gives better predictions across all parameters. We see a familiar pattern emerge in the 2D CNNs; they have difficulty predicting spacecraft impact parameter and more often predict chirality incorrectly. This in turn leads to greater inaccuracies in $\phi$ and $\theta$ predictions. Given that the 1D CNN out performed the 2D CNNs in both training experiments, we focus only on the 1D architecture when evaluating network performance on actual spacecraft measurements.

\subsection{Application to Wind Catalog Flux Ropes}
\begin{figure}[H]
  \begin{center}
  \includegraphics[width=\textwidth]{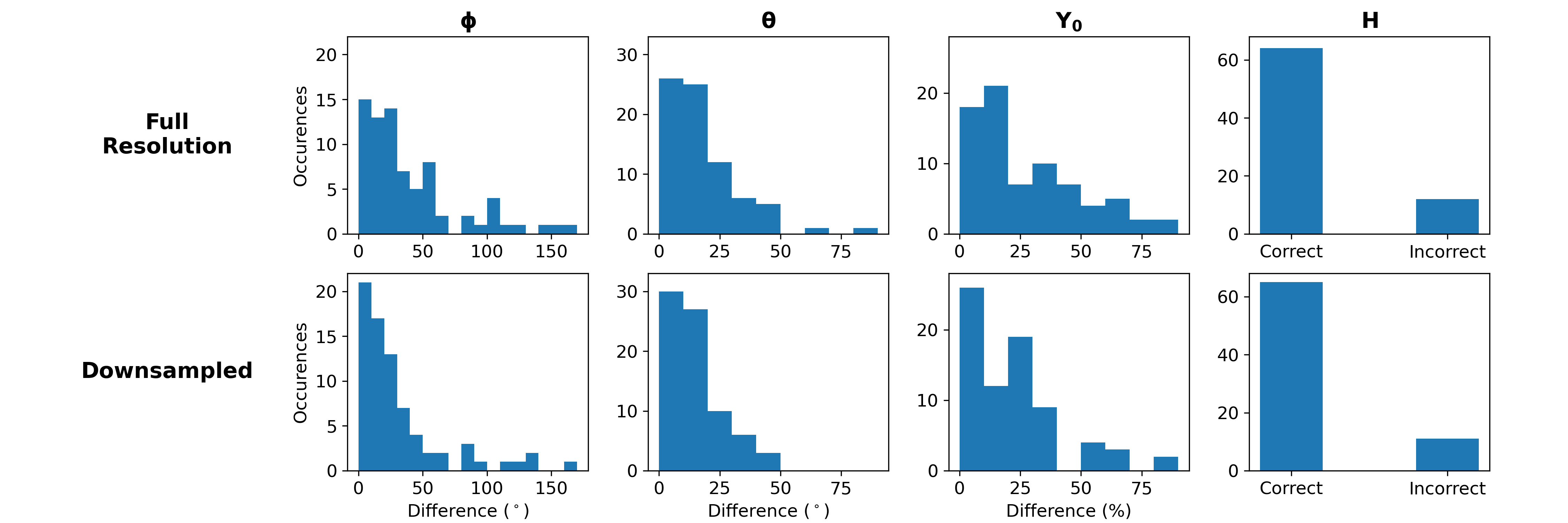}
  \end{center}
  \caption{The 1D CNN parameter prediction error for the Wind event test set of full duration flux rope crossings. Two sets of predictions were generated: One from processing 1 minute Wind MFI measurements and the second from processing the Wind MFI measurements down-sampled to the 15 minute averages. The human-fit parameter values from the published ICME catalog were compared against neural network predictions. The overall error magnitude is greater than when tested on synthetic input but shows the same trend. Predictions improve when the down-sampled input is used. }
  \label{fig:Wind_Full_1}
\end{figure}

To assess the transfer-ability of this technique to real-time use, we applied the 1D CNN trained on full duration flux ropes to the 75 selected Wind events described in Section  \ref{Wind_data} with the data processed in two ways.  The first approach, which we label Full Resolution, is where we simply use the window smoothing before interpolating the event down to 50 points.  The second approach, called Downsampled, first applies 15 minute averaging before smoothing and interpolation. The idea being that the Downsampled approach would further reduce fluctuations found inside Wind flux ropes. Comparing Full Resolution and Downsampled would help us isolate the impacts of fluctuations. The difference histograms in Figure \ref{fig:Wind_Full_1} show the result of comparing the fit parameters from \citet{Nieves_2019}(N-C) with the neural network predicted parameters.  We note that our neural network was trained with force free synthetic flux ropes ($C_{10}$ parameter equal to 1). The N-C fittings allowed for deviations from a force free flux rope. This difference likely played a role in the discrepancies between neural network predictions and the human expert's fits.

Most $\phi$ predictions were within $50^\circ$ of the hand-fit value but the maximum error was over $150^\circ$.  The $\theta$ errors tend to be less than $25^\circ$ with a maximum around $80^\circ$.  Most $Y_0$ predictions are within $30\%$ of the comparison values with with a maximum near $80\%$.  Across all these real-valued key parameters, the predictions made from the Downsampled input display a less skewed error distribution with a higher percentage of the predictions having relatively small error.  The network produced similar results predicting chriality ($H$) when fed with Full Resolution and Downsampled input.

\begin{figure}[H]
  \begin{center}
  \includegraphics[width=\textwidth]{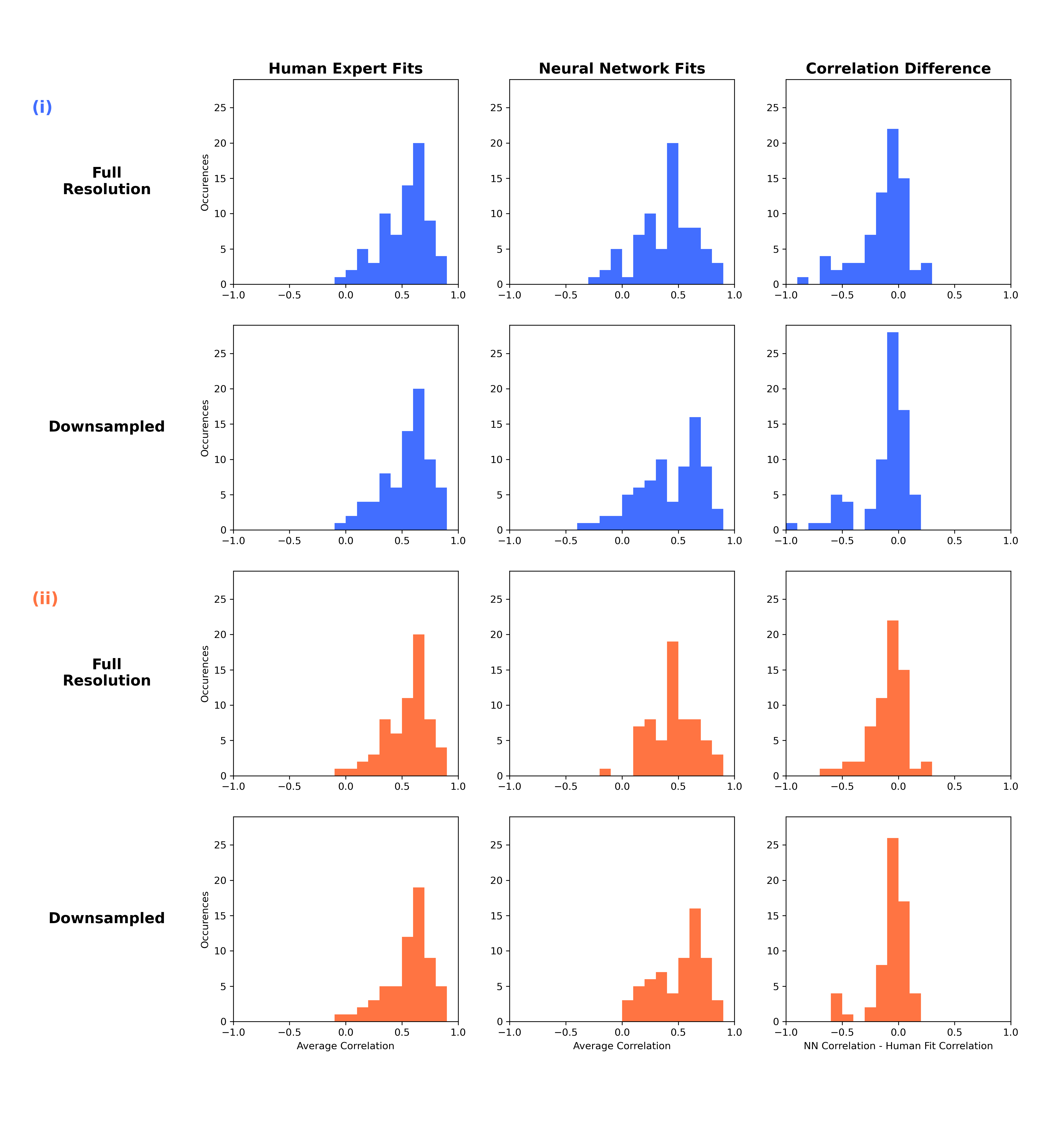}
  \end{center}
  \caption{Correlation distributions and comparisons for the Wind event test set of full duration flux rope crossings. Column 1 shows the correlation score between the human-fit parameters and the Wind measurements.  Column 2 shows the correlation between the 1D CNN prediction and the Wind measurements.  The third column displays the difference between the human-fit correlations and the CNN prediction correlations. (i) Includes all 75 Wind test cases. (ii) Includes cases where the CNN $Y_0$ prediction matched human expert's chirality only. When the CNN predicts $Y_0$ correctly, the correlation to Wind data is similar to that of the human-expert.}
  \label{fig:chi_comparison}
\end{figure}

We extend this comparison with analysis of average correlation coefficient. We display correlation between the interpolated Wind observations and the magnetic field vectors generated using the N-C fit parameters as well as the the correlation between interpolated Wind observations and magnetic field vectors generated from neural network predictions. Figure \ref{fig:chi_comparison} column 1 shows the distribution of average correlation between the human-fit model and Wind data. Column 2 is the distribution of average correlation between the CNN fit and Wind data. Displayed in column 3 is the difference histogram showing the neural network correlation minus the hand-fit correlation for each of the Wind flux rope events. Positive values indicate the neural network produced a statistically more reliable fit.Panel \ref{fig:chi_comparison}(i) shows these distributions for all of the 75 events. Panel \ref{fig:chi_comparison}(ii) shows the distributions when we consider only the events in which the CNN predicted the same chirality as the human-fit. We see good agreement in average correlation coefficients when the predictions are used to reconstruct the magnetic field time series. The shape of the distributions are similar to those from the comparison with human expert fits and an event by event comparison with human expert fits leads to a difference histogram nearly centered at zero. When we look at the Downsampled neural network predictions with chirality prediction matching the chirality of the human expert (Fig. \ref{fig:chi_comparison} (ii)) we see no negative correlations.

We next applied the network trained on full and partial duration flux ropes to the aforementioned subset of 75 Wind Fr events. A summary of the results are shown in Table \ref{Table:partial_stats_summary} where we list median differences between network predictions and hand-fit values as a function of flux rope observed. Also shown are the percentage of events where predicted chirality and hand-fit chirality match. The median difference in longitude ranges from $58^\circ$ to $89^\circ$; in latitude from $31^\circ$ to $50^\circ$; and in impact parameter from $36\%$ to $53\%$.  The network predicted the chirality correctly between $52\%$ and $65\%$ of the time.

\begin{table}[H]
\begin{center}
\begin{tabular}{|c|c|c|c|c|c|c|}
\hline
  \% Observed & $\phi$ & $\theta$ & $Y_0$ & $H$ 
    \\ \hline
10  & $89^{\circ}$ & $50^{\circ}$ & 36\% & 63\% \\ \hline
20  & $66^{\circ}$ & $42^{\circ}$ & 39\% & 52\%  \\ \hline
30  & $69^{\circ}$ & $32^{\circ}$ & 53\% & 60\% \\ \hline
40  & $64^{\circ}$ & $37^{\circ}$ & 39\% & 60\% \\ \hline
50  & $70^{\circ}$ & $33^{\circ}$ & 37\% & 60\% \\ \hline
60  & $69^{\circ}$ & $37^{\circ}$ & 51\% & 60\% \\ \hline
70  & $73^{\circ}$ & $31^{\circ}$ & 44\% & 64\% \\ \hline
80  & $73^{\circ}$ & $34^{\circ}$ & 53\% & 65\% \\ \hline
90  & $73^{\circ}$ & $33^{\circ}$ & 47\% & 56\% \\ \hline
100 & $58^{\circ}$ & $42^{\circ}$ & 44\% & 60\% \\ \hline

\end{tabular}
\end{center}
\caption{Wind event summary statistics as a function of percentage of flux rope observed for the 1D network trained with both full and partial duration flux ropes. Human-fit parameters are compared to neural network predictions and the $\phi$, $\theta$, and $Y_0$ columns are median differences between the two. The $H$ column is the percentage of events where the chirality prediction matches hand-fit value.}
\label{Table:partial_stats_summary}
\end{table}

\subsection{Number of Wind Events to Train a Network}

Experimenting with synthetic and real flux ropes raised an interesting question: How many real flux ropes are needed to train a neural network and how many suitable flux ropes are available for such a study? We can not answer this question conclusively. As discussed in a previous section, and elaborated on below, neural networks trained on synthetic events do not transfer perfectly to Wind. However, we can perform one additional experiment to roughly gauge an answer. 

We re-used the train-validation-test split of our synthetic flux ropes mentioned in Section 2.2. We then set up a loop of nine iterations. In each iteration, we randomly selected a diminishing subset of the training data, trained a 1D CNN with that subset, and then evaluated the trained model on the testing set of 19,600 synthetic flux ropes. The subsets were selected at random to simulate what happens in practice where we cannot dictate the orientation of flux ropes observed by a spacecraft. The testing consisted of using the trained CNN to make orientation predictions for each of the 19,600 test flux ropes, use those orientation predictions to create the corresponding magnetic field profiles, and correlate those magnetic field profiles with the magnetic field profiles of the test flux rope. As an evaluation metric, we computed the percentage of the test correlations greater than or equal to 0.75. Within each iteration, the subsetting-training-prediction-correlation workflow was repeated three times to investigate how the random sub-setting might impact the results.

Table \ref{Table:num_synthetic} lists the results. Over ninety percent of testing events have an average correlation coefficient above 0.75 as long as the training set size is over 200 events. Put another way, a 1D CNN trained with roughly 200 events produces average correlation coefficients on par with the 2D 3-input CNN (middle panel of \ref{fig:Results_3_full}). We do note, however, that our experiment is based on training the network with a specific flux rope model and simulated (synthetic) flux ropes. The specific flux rope model chosen will play a role as more complex descriptions of flux ropes (i.e. taking into account compression/expansion) will have more output parameters, which in turn will impact accuracy. In addition, these synthetic flux ropes do not take into account the turbulent fluctuations found in real flux ropes - a further source of prediction error. Nevertheless, it is interesting to note that we may be tantalizingly close to a neural network trained on real observations. There are 151 Wind events in \citet{Nieves_2019} that could potentially be used in training. The HELIO4CAST ICME catalog version 2.1 \citep{Moestl2020} has over 1,000 ICMEs identified from multiple spacecraft. It is unknown how many of these ICMEs have associated flux ropes. Once identified, assuming there are enough, those flux ropes will need to be fit by human experts to provide labeled data for supervised learning. Nevertheless, our experiment provides the intriguing result that a few hundred more events may be all that is needed. Existing ICME catalogs may hold enough events that a concerted effort could lead to training set of real flux ropes in the coming years.

\begin{table}[H]
\begin{center}
\begin{tabular}{|c|c|c|}
\hline
  \# Flux Ropes in Training & $\% \geq 0.75$ & SD    \\ \hline
29440  & 99 & 0.5 \\ \hline
14592  & 99 & 0.4 \\ \hline
7168  & 98 & 0.05 \\ \hline
3584  & 97 & 0.4  \\ \hline
1792  & 97 & 1.3  \\ \hline
1024  & 97 & 0.6  \\ \hline
512  & 95 & 0.3   \\ \hline
256  & 93 & 0.6   \\ \hline
128  & 88 & 0.88   \\ \hline

\end{tabular}
\end{center}
\caption{Percentage of synthetic flux rope predictions with an average correlation coefficient of 0.75 or greater as a function of training set size. The 1D CNN was used for training. Each training set size was repeated three time, each time taking a different random sample. The percentages reported are the average of the three repetitions. The SD column lists the standard deviation of the three repetitions.}
\label{Table:num_synthetic}
\end{table}

\section{Discussion and Conclusions}

Our experiments have demonstrated that convolutional neural networks are capable of providing extremely reliable characterizations of flux ropes from synthetic data. A trained network can use the structure of simulated magnetic field vectors to learn filters that map to accurate flux rope key parameter predictions; successfully inferring large scale, 3D information from single-point measurements.

When trained only on examples of full duration flux ropes, all three architectures predict key parameters of a flux rope which correlate well with the input data; however the best performing is the 1D network that feeds on time series data. Although the 2D networks that use hodogram style input do not see the same, perfect accuracy in predicting the chirality as the 1D network, the difference is statistically minor. The biggest weakness in the hodogram-input CNNs is when interpreting flux rope traces generated with a high spacecraft impact parameter. It is possible that similarities in hodogram shape profile between low- and high-valued $Y_0$ are activating similar filters in the 2D networks and leading to poor predictions in these cases.

When we extend the synthetically trained networks to include both partial and full duration traces through flux ropes, we still find this approach highly accurate. The CNNs are capable of making reliable predictions having only seen a fraction of the full flux rope.  Although the overall discrepancy between the true and predicted values is higher than when done with only full duration traces, all median differences are well within a tolerable limit.  In these idealized, synthetic, circular-cylindrical flux ropes even the poorest performing network is able to predict orientation angles with a median error under $8^\circ$ after only observing $10\%$ of a simulated spacecraft crossing.  The 1D network here, at only $10\%$ observed, is able to give predictions with lower median difference than the 2D networks do when trained and tested with only full flux ropes.  All three models show a peak performance at some point prior to seeing $100\%$ of the flux rope crossing, perhaps due to some similarity in shape between low percentage of observation and high percentage.  It is interesting to note that the 2D networks hit their peak predictive point earlier than the 1D CNN, 2D one input at $50-60\%$ and 2D three input even earlier at $40-50\%$. This suggests that research into where the convolutional network is looking (for example, with the Grad-CAM method \cite{grad_cam}) can help us further understand the benefits and limitations of hodograms and time series as inputs. Future research will examine where the network is focusing its attention and if this can be exploited for more accurate predictions earlier in the forecasting process.

With the success of the 1D CNN in real-time forecasting from idealized synthetic data, we evaluated this trained 1D network on partial Wind event data. Overall, the neural network struggles to reproduce the accuracy achieved on the synthetic data set. Unlike the synthetic case, we see no trend towards a peak performance point dependent on the amount of flux rope observed.  When looked at on a case-by-case basis, there are a few specific events in which the neural network is able to make accurate predictions after only seeing a fraction of the flux rope. In general, however, the median difference in angle and impact parameter prediction falls well outside any tolerance levels for useful prediction and the chirality is only correct approximately $60\%$ of the time. Clearly, the partial-trained CNN cannot be transferred as-is to real-time application, but insight can be found by examining the results of the full duration network evaluated with Wind events.

Applying the 1D CNN trained only on full duration synthetic flux ropes to in situ Wind events, we again see the individual parameter predictions show significant deviation from hand fit values. However, we note lower median differences and higher $H$ accuracy than when the network trained on both full and partial events was applied to Wind. Using down-sampled input improves this even further.
Yet, by looking at the average correlation scores we see that the flux rope analytical model is robust to small deviations - small changes in longitude in particular do not lead to significant differences in reconstructed time series. We also find the neural network robust to variation in solar wind speed, expansion/compression, duration, and to some degree, magnetic field fluctuations. The neural networks were trained on synthetic data that was all generated with a simulated solar wind velocity of 450 km/s and simulated flux rope radius of 0.07 AU; yet, are able to offer reasonable predictions for Wind Fr events having significant differences in solar wind speed, expansion/compression, duration, and magnetic field fluctuations.

The neural network gives reliable predictions in a number of events and exhibits a distribution of average correlations that is qualitatively similar to those from the human expert. As evident in the right-most column of Figure \ref{fig:chi_comparison}, the neural network results in better average correlation in nearly half of the 75 events. When we consider only cases in which the network prediction for $H$ matches the human-fit $H$ the correlation to Wind data is even greater. 

\begin{figure}[H]
  \begin{center}
  \includegraphics[width=\textwidth]{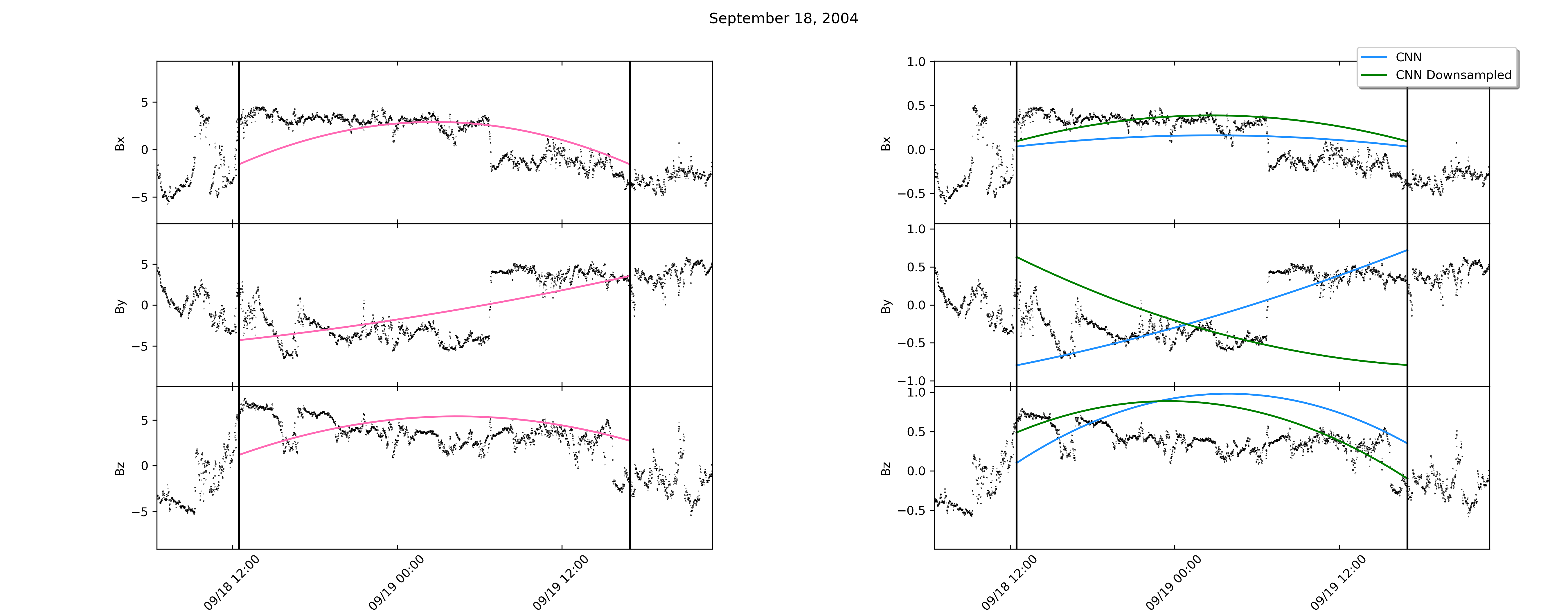}
  \end{center}
  \caption{Wind MFI measurements for September 18 - 19, 2004 overlaid with simulated spacecraft flux rope crossings. The left panel shows the crossing described by the human-fit parameters.  The right panel shows the crossings described by the CNN predictions when it is given both the Full and Downsampled input.}
  \label{fig:Wind_Example_comp}
\end{figure}

Analysis reveals two primary reasons the neural network performs less accurately on Wind events; incorrect physical model (Wind flux ropes not fitting the circular cylindrical assumptions) and internal physical processes (such as fluctuations and discontinuities) that alter the expected magnetic field profile of a smooth flux rope. An example of a flux rope with magnetic field fluctuations and a discontinuity is shown in the event with MO beginning on September 18, 2004 in Figure \ref{fig:Wind_Example_comp}. Down-sampling the Wind magnetic field data from 1-minute to 15-minutes prior to interpolating to 50 points reduces the difference between neural network predictions and hand-fit values. The down-sampling further smooths out the magnetic field time series removing small-scale fluctuations. However, down-sampling cannot account for all observed internal physical processes that lead to a deviation from the expected smooth flux rope profile. The September 2004 event illustrates how differences in data processing can have a strong effect on the resulting prediction. In this particular example, the predictions made from the Wind data without prior averaging match the hand fit predictions well, while those from the down-sampled input clearly lost important information. The choice of 15-minute averaging was arbitrary and is presented here to highlight how data pre-processing can have both positive and negative impacts on prediction accuracy. It remains for future research to systematically address fluctuations and determine an optimal input resolution. 

Of the total 151 Wind Fr events in \citet{Nieves_2019}, only 41\% were classified as a flux rope by the neural network developed in \citet{dossantos_2020} when trained with no fluctuations. This same network classified 84\% and 76\% as flux ropes when trained with synthetic data augmented with 5\% and 10\% Gaussian fluctuations, respectively. In other words, some of the Wind events on which we do poorly finding good parametrization, would not have been considered a flux rope by the first step of an automated fitting workflow. At present, magnetic field fluctuations are not fully accounted for in flux rope analytical models and pre-processing of neural network input data does not fully address the discrepancy between synthetic and spacecraft observed flux ropes. Accurately accounting for fluctuations in measured data appears to be a significant factor for improving an automated space weather forecasting pipeline. Early experimentation with 5\% Gaussian fluctuations in our study did not lead to significant improvement. Solar wind and flux rope turbulence is known to be non-Gaussian. Yet, at present, a complete understanding of turbulence leads analytical models lacking in this regard. We choose to not introduce non-realistic fluctuations and instead will explore physics-based turbulence enhancements to the analytical model in future research.  

The ultimate source of prediction error in any CNN is in the inputs not matching any of the learned filters. In the case of Wind events, we notice that the neural network trained with only full duration flux ropes incorrectly predicts chirality in nearly 20\% of Wind events. This leads to poor correlation coefficients as the reconstructed time series do not match the Wind observations. Yet, across all implementations of the CNNs with synthetic data the CNNs overwhelmingly identify the correct chirality. This indicates that the convolutional filters the network learned to predict chirality do not transfer to Wind events; that the filters learned to focus on a quality in the synthetic data that is not shared in the real observations. 
Interestingly, down-sampling has no effect on chirality predictions. We believe this source of error is related to the physical model chosen to simulate the flux ropes. Wind flux ropes show deviations from the circular cylindrical assumption. This opens the door to tantalizing future evaluations of physics-based flux rope models using an ensemble of neural networks, each trained with a different physical model.

Partial duration predictions and real-time forecasting are not really feasible at this time due in large part to features in the real data that are not present in the training set, though the concept of using CNNs to infer 3D geometric parameters from an in situ measurement have been borne out. Additionally, the neural networks have helped highlight the limitations of the physics-based model and even suggested better fittings of some Wind flux ropes. Future work will include implementing a single, physics-based loss function into the CNN to replace the four separate loss functions in the current design as well as enhancing analytical flux rope models to produce training data that includes more realistic turbulence and asymmetry.

\section*{Conflict of Interest Statement}
Author Ayris Narock is employed by ADNET Systems Inc. and is contracted to NASA/Goddard Space Flight Center to carry out scientific software development and data analysis support. Despite the affiliation with a commercial entity, author Ayris Narock has no relationships that could be construed as a conflict of interest related to this work. The remaining authors declare that the research was conducted in the absence of any commercial or financial relationships that could be construed as a potential conflict of interest.

\section*{Author Contributions}
The authors confirm contribution to the paper as follows: study conception and design: TN, AN, LFGS, and TN-C; flux rope analytical model development and coding: TN-C developed the analytical model and initial version of the code. AN ported the code to language/version used in this study; synthetic data generation: TN; neural network design and implementation; TN, AN, LFGS; neural network training and testing: TN; analysis and interpretation of results: TN, AN, LFGS, and TN-C; draft manuscript preparation: TN, AN, LFGS, and TN-C;. All authors reviewed the results and approved the final version of the manuscript. 

\section*{Funding}
L.F.G.S was supported by NASA Grant 80NSSC20K1580

\section*{Acknowledgments}
The first author would like to acknowledge Ron Lepping, Daniel Berdichevsky, and Chin-Chun Wu for their many helpful discussions surrounding flux rope physics during earlier projects involving flux rope detection and analysis.



\bibliographystyle{frontiersinHLTH&FPHY} 
\bibliography{paper}

\begin{thebibliography}{29}
\expandafter\ifx\csname natexlab\endcsname\relax\def\natexlab#1{#1}\fi
\expandafter\ifx\csname urlstyle\endcsname\relax
  \expandafter\ifx\csname doi\endcsname\relax
  \def\doi#1{doi:\discretionary{}{}{}#1}\fi \else
  \expandafter\ifx\csname doi\endcsname\relax
  \def\doi{doi:\discretionary{}{}{}\begingroup \urlstyle{rm}\Url}\fi \fi
\expandafter\ifx\csname selectlanguage\endcsname\relax
  \def\selectlanguage#1{}\fi

\bibitem[{Baker and Lanzerotti(2008)}]{baker_2008}
Baker DN, Lanzerotti LJ.
\newblock A continuous l1 presence required for space weather.
\newblock {\em Space Weather\/} {\bf 6} (2008).
\newblock \doi{https://doi.org/10.1029/2008SW000445}.

\bibitem[{Kilpua et~al.(2017{\natexlab{a}})Kilpua, Koskinen, and
  Pulkkinen}]{Kilpua2017}
Kilpua E, Koskinen HEJ, Pulkkinen TI.
\newblock Coronal mass ejections and their sheath regions in interplanetary
  space.
\newblock {\em Living Reviews in Solar Physics\/} {\bf 14}
  (2017{\natexlab{a}}).
\newblock \doi{https://doi.org/10.1007/s41116-017-0009-6}.

\bibitem[{Burlaga(1988)}]{burlaga_1988}
Burlaga LF.
\newblock Period doubling in the outer heliosphere.
\newblock {\em Journal of Geophysical Research: Space Physics\/} {\bf 93}
  (1988) 4103--4106.
\newblock \doi{https://doi.org/10.1029/JA093iA05p04103}.

\bibitem[{Burlaga et~al.(1981)Burlaga, Sittler, Mariani, and
  Schwenn}]{burlaga_1981}
Burlaga L, Sittler E, Mariani F, Schwenn R.
\newblock Magnetic loop behind an interplanetary shock: Voyager, helios, and
  imp 8 observations.
\newblock {\em Journal of Geophysical Research: Space Physics\/} {\bf 86}
  (1981) 6673--6684.
\newblock \doi{https://doi.org/10.1029/JA086iA08p06673}.

\bibitem[{Klein and Burlaga(1982)}]{klein_burlaga_1982}
Klein LW, Burlaga LF.
\newblock Interplanetary magnetic clouds at 1 au.
\newblock {\em Journal of Geophysical Research: Space Physics\/} {\bf 87}
  (1982) 613--624.
\newblock \doi{https://doi.org/10.1029/JA087iA02p00613}.

\bibitem[{{Vourlidas}(2014)}]{Vourlidas_2014}
{Vourlidas} A.
\newblock {The flux rope nature of coronal mass ejections}.
\newblock {\em Plasma Phys. Contr. Fus.\/} {\bf 56} (2014) 064001.
\newblock \doi{10.1088/0741-3335/56/6/064001}.

\bibitem[{Jian et~al.(2006)Jian, Russell, Luhmann, and Skoug}]{Jian2006}
Jian L, Russell CT, Luhmann JG, Skoug RM.
\newblock Properties of interplanetary coronal mass ejections at one au during
  1995 2004.
\newblock {\em Solar Physics\/} {\bf 239} (2006) 393--436.
\newblock \doi{https://doi.org/10.1007/s11207-006-0133-2}.

\bibitem[{Manchester et~al.(2017)Manchester, Kilpua, Liu, Lugaz, Riley, P.
  et~al.}]{manchester_2017}
Manchester W, Kilpua E, Liu Y, Lugaz N, Riley P, P T Torok, et~al.
\newblock The physical processes of cme/icme evolution.
\newblock {\em Space Sci. Rev.\/} {\bf 212(3-4), 1159} (2017).

\bibitem[{{Lepping} et~al.(1990){Lepping}, {Jones}, and
  {Burlaga}}]{Lepping_Jones_Burlaga_1990}
{Lepping} RP, {Jones} JA, {Burlaga} LF.
\newblock {Magnetic field structure of interplanetary magnetic clouds at 1 AU}.
\newblock {\em Journal of Geophysical Research\/} {\bf 95} (1990) 11957--11965.
\newblock \doi{10.1029/JA095iA08p11957}.

\bibitem[{{Nieves-Chinchilla} et~al.(2019){Nieves-Chinchilla}, {Jian},
  {Balmaceda}, {Vourlidas}, {dos Santos}, and {Szabo}}]{Nieves_2019}
{Nieves-Chinchilla} T, {Jian} LK, {Balmaceda} L, {Vourlidas} A, {dos Santos}
  LFG, {Szabo} A.
\newblock {Unraveling the Internal Magnetic Field Structure of the
  Earth-directed Interplanetary Coronal Mass Ejections During 1995 - 2015}.
\newblock {\em Solar Physics\/} {\bf 294} (2019) 89.
\newblock \doi{10.1007/s11207-019-1477-8}.

\bibitem[{{Nieves-Chinchilla} et~al.(2018){Nieves-Chinchilla}, {Vourlidas},
  {Raymond}, {Linton}, {Al-haddad}, {Savani} et~al.}]{Nieves_2018}
{Nieves-Chinchilla} T, {Vourlidas} A, {Raymond} JC, {Linton} MG, {Al-haddad} N,
  {Savani} NP, et~al.
\newblock {Understanding the Internal Magnetic Field Configurations of ICMEs
  Using More than 20 Years of Wind Observations}.
\newblock {\em Solar Physics\/} {\bf 293} (2018) 25.
\newblock \doi{10.1007/s11207-018-1247-z}.

\bibitem[{Rodr{\'\i}guez-Garc{\'\i}a et~al.(2021)Rodr{\'\i}guez-Garc{\'\i}a,
  G{\'o}mez-Herrero, Zouganelis, Balmaceda, Nieves-Chinchilla, Dresing
  et~al.}]{rodriguez_2021}
Rodr{\'\i}guez-Garc{\'\i}a L, G{\'o}mez-Herrero R, Zouganelis I, Balmaceda L,
  Nieves-Chinchilla T, Dresing N, et~al.
\newblock The unusual widespread solar energetic particle event on 2013 august
  19: Solar origin and particle longitudinal distribution.
\newblock {\em Astronomy \& Astrophysics\/} {\bf 653} (2021) A137.

\bibitem[{Kilpua et~al.(2017{\natexlab{b}})Kilpua, Koskinen, and
  Pulkkinen}]{kilpua_2017}
Kilpua E, Koskinen HE, Pulkkinen TI.
\newblock Coronal mass ejections and their sheath regions in interplanetary
  space.
\newblock {\em Living Reviews in Solar Physics\/} {\bf 14} (2017{\natexlab{b}})
  1--83.

\bibitem[{Gosling et~al.(1973)Gosling, Pizzo, and Bame}]{gosling_1973}
Gosling JT, Pizzo V, Bame SJ.
\newblock Anomalously low proton temperatures in the solar wind following
  interplanetary shock waves—evidence for magnetic bottles?
\newblock {\em Journal of Geophysical Research (1896-1977)\/} {\bf 78} (1973)
  2001--2009.
\newblock \doi{https://doi.org/10.1029/JA078i013p02001}.

\bibitem[{{Camporeale}(2019)}]{Camporeale_2019}
{Camporeale} E.
\newblock {The Challenge of Machine Learning in Space Weather: Nowcasting and
  Forecasting}.
\newblock {\em Space Weather\/} {\bf 17} (2019) 1166--1207.
\newblock \doi{10.1029/2018SW002061}.

\bibitem[{Nguyen et~al.(2018)Nguyen, Fontaine, Aunai, Vandenbossche, Jeandet,
  Lemaitre et~al.}]{Nguyen2018}
Nguyen G, Fontaine D, Aunai N, Vandenbossche J, Jeandet A, Lemaitre G, et~al.
\newblock Machine learning methods to identify icmes automatically.
\newblock {\em 20th EGU General Assembly, EGU2018, Proceedings from the
  conference held 4-13 April, 2018 in Vienna, Austria\/}  (2018) 1963.

\bibitem[{{dos Santos} et~al.(2020){dos Santos}, {Narock}, {Nieves-Chinchilla},
  {Nu{\~n}ez}, and {Kirk}}]{dossantos_2020}
{dos Santos} LFG, {Narock} A, {Nieves-Chinchilla} T, {Nu{\~n}ez} M, {Kirk} M.
\newblock {Identifying Flux Rope Signatures Using a Deep Neural Network}.
\newblock {\em Solar Physics\/} {\bf 295} (2020) 131.
\newblock \doi{10.1007/s11207-020-01697-x}.

\bibitem[{Reiss et~al.(2021)Reiss, Möstl, Bailey, Rüdisser, Amerstorfer,
  Amerstorfer et~al.}]{Reiss2021}
Reiss MA, Möstl C, Bailey RL, Rüdisser HT, Amerstorfer UV, Amerstorfer T,
  et~al.
\newblock Machine learning for predicting the bz magnetic field component from
  upstream in situ observations of solar coronal mass ejections.
\newblock {\em Space Weather\/} {\bf 19} (2021).
\newblock \doi{https://doi.org/10.1029/2021SW002859}.

\bibitem[{{Nieves-Chinchilla} et~al.(2016){Nieves-Chinchilla}, {Linton},
  {Hidalgo}, {Vourlidas}, {Savani}, {Szabo} et~al.}]{Nieves_C_2016}
{Nieves-Chinchilla} T, {Linton} MG, {Hidalgo} MA, {Vourlidas} A, {Savani} NP,
  {Szabo} A, et~al.
\newblock {A Circular-cylindrical Flux-rope Analytical Model for Magnetic
  Clouds}.
\newblock {\em The Astrophysical Journal\/} {\bf 823} (2016) 27.
\newblock \doi{10.3847/0004-637X/823/1/27}.

\bibitem[{LeCun and Bengio(1995)}]{lecun1995}
LeCun Y, Bengio Y.
\newblock Convolutional networks for images, speech, and time series.
\newblock {\em The Handbook of Brain Theory and Neural Networks\/} {\bf 3361}
  (1995) 1995.

\bibitem[{{Kingma} and {Ba}(2015)}]{kingma_2015}
{Kingma} D, {Ba} J.
\newblock Adam: A method for stochastic optimization.
\newblock {\em Bengio, Y., LeCun, Y. (eds.) 3rd International Conference on
  Learning Representations, ICLR 2015, San Diego, CA, USA, May 7-9, 2015,
  Conference Track Proceedings\/}  (2015).

\bibitem[{Chollet et~al.(2015)}]{keras}
[Dataset] Chollet F, et~al.
\newblock Keras.
\newblock \url{https://keras.io} (2015).

\bibitem[{Abadi et~al.(2015)Abadi, Agarwal, Barham, Brevdo, Chen, Citro
  et~al.}]{tensorflow}
[Dataset] Abadi M, Agarwal A, Barham P, Brevdo E, Chen Z, Citro C, et~al.
\newblock {TensorFlow}: Large-scale machine learning on heterogeneous systems
  (2015).
\newblock Software available from tensorflow.org.

\bibitem[{Harris et~al.(2020)Harris, Millman, van~der Walt, Gommers, Virtanen,
  Cournapeau et~al.}]{numpy}
Harris CR, Millman KJ, van~der Walt SJ, Gommers R, Virtanen P, Cournapeau D,
  et~al.
\newblock Array programming with {NumPy}.
\newblock {\em Nature\/} {\bf 585} (2020) 357--362.
\newblock \doi{10.1038/s41586-020-2649-2}.

\bibitem[{Virtanen et~al.(2020)Virtanen, Gommers, Oliphant, Haberland, Reddy,
  Cournapeau et~al.}]{scipy}
Virtanen P, Gommers R, Oliphant TE, Haberland M, Reddy T, Cournapeau D, et~al.
\newblock {{SciPy} 1.0: Fundamental Algorithms for Scientific Computing in
  Python}.
\newblock {\em Nature Methods\/} {\bf 17} (2020) 261--272.
\newblock \doi{10.1038/s41592-019-0686-2}.

\bibitem[{{Lepping} et~al.(1995){Lepping}, {Ac{\~{u}}na}, {Burlaga}, {Farrell},
  {Slavin}, {Schatten} et~al.}]{Lepping_1995}
{Lepping} RP, {Ac{\~{u}}na} MH, {Burlaga} LF, {Farrell} WM, {Slavin} JA,
  {Schatten} KH, et~al.
\newblock {The Wind Magnetic Field Investigation}.
\newblock {\em Space Science Reviews\/} {\bf 71} (1995) 207--229.
\newblock \doi{10.1007/BF00751330}.

\bibitem[{{Ogilvie} et~al.(1995){Ogilvie}, {Chornay}, {Fritzenreiter},
  {Hunsaker}, {Keller}, {Lobell} et~al.}]{Ogilvie_1995}
{Ogilvie} KW, {Chornay} DJ, {Fritzenreiter} RJ, {Hunsaker} F, {Keller} J,
  {Lobell} J, et~al.
\newblock {SWE, A Comprehensive Plasma Instrument for the Wind Spacecraft}.
\newblock {\em Space Science Reviews\/} {\bf 71} (1995) 55--77.
\newblock \doi{10.1007/BF00751326}.

\bibitem[{Moestl et~al.(2020)Moestl, Weiss, Bailey, and Reiss}]{Moestl2020}
Moestl C, Weiss A, Bailey R, Reiss M.
\newblock Helio4cast interplanetary coronal mass ejection catalog v2.1.
\newblock {\em figshare. Dataset.\/}  (2020).
\newblock \doi{https://doi.org/10.6084/m9.figshare.6356420.v11}.

\bibitem[{Selvaraju et~al.(2017)Selvaraju, Cogswell, Das, Vedantam, Parikh, and
  Batra}]{grad_cam}
Selvaraju RR, Cogswell M, Das A, Vedantam R, Parikh D, Batra D.
\newblock Grad-cam: Visual explanations from deep networks via gradient-based
  localization.
\newblock {\em 2017 IEEE International Conference on Computer Vision (ICCV)\/}
  (2017), 618--626.
\newblock \doi{10.1109/ICCV.2017.74}.

\end{thebibliography}



\end{document}